\documentclass[12pt,preprint,landscape]{aastex}
\usepackage{natbib}
\usepackage{amsmath}
\usepackage{graphicx}
\usepackage{lscape}
\usepackage{epsfig}

\def\Dwa{$\,$\uppercase\expandafter{\romannumeral5}$\,$}

\def\sless{\lower2pt\hbox{$\buildrel {\scriptstyle <}
   \over {\scriptstyle\sim}$}}

\def\sgreat{\lower2pt\hbox{$\buildrel {\scriptstyle >}
   \over {\scriptstyle\sim}$}}

\begin{document}


\title{Expectations for the Hard X-ray Continuum and Gamma-ray Line Fluxes from the 
Type Ia supernova SN 2014J in M82}

\author{Lih-Sin The\altaffilmark{1}, Adam Burrows\altaffilmark{2}}   
\altaffiltext{1}{Department of Physics and Astronomy, Clemson University, 
SC 29634 USA; tlihsin@clemson.edu}
\altaffiltext{2}{Department of Astrophysical Sciences, Princeton University, 
Princeton, NJ 08544 USA; burrows@astro.princeton.edu}

\begin{abstract}
The hard X-ray continuum and gamma-ray lines from a Type Ia supernova dominate 
its integrated photon emissions and can provide unique diagnostics
of the mass of the ejecta, the $^{56}$Ni yield and spatial distribution, its kinetic energy 
and expansion speed, and the mechanism of explosion. Such signatures and their time behavior 
``X-ray" the bulk debris field in direct fashion, and do not depend upon the ofttimes problematic
and elaborate UV, optical, and near-infrared spectroscopy and radiative transfer that have 
informed the study of these events for decades.  However, to date no hard photons have ever
been detected from a Type Ia supernova in explosion. With the advent of the supernova SN 2014J in M82,
at a distance of $\sim$3.5 Mpc, this situation may soon change.  Both NuSTAR and INTEGRAL have 
the potential to detect SN 2014J, and, if spectra and light curves can be measured, would usefully
constrain the various explosion models published during the last $\sim$thirty years.  In 
support of these observational campaigns, we provide predictions for the hard X-ray 
continuum and gamma-line emissions for fifteen Type Ia explosion models gleaned 
from the literature.  The model set, containing as it does deflagration, delayed detonation, 
merger detonation, pulsational delayed detonation, and sub-Chandrasekhar helium detonation models, collectively 
spans a wide range of properties and, hence, signatures. We provide a brief discussion 
of various diagnostics (with examples), but importantly make the  spectral and line results 
available electronically to aid in the interpretation of the anticipated data. 
\end{abstract}


\section{Introduction}
\label{intro}

A Type Ia supernova explosion is best modeled as a thermonuclear 
explosion (of carbon and oxygen, and perhaps some helium) 
of a degenerate object near one solar mass (M$_{\odot}$), with
a kinetic energy near 10$^{51}$ ergs and the production of 
$\sim$0.5 to $\sim$1.0 M$_{\odot}$ of radioactive $^{56}$Ni (Pankey 1962; 
Colgate \& McKee 1969; Clayton, Colgate, \& Fishman 1969).  The object could be 
a Chandrasekhar white dwarf, the merger product of two white dwarfs \citep{1993A&A...270..223K,1996ApJ...472L..81H}, 
or a slightly sub-Chandrasekhar white dwarf.  The explosions could be subsonic deflagrations \citep{w7},
sub-Chandrasekhar detonations \citep{1996ApJ...472L..81H}, delayed detonations 
\citep{WW91,1992ApJ...393L..55Y,1995ApJ...443...89H,1998ApJ...495..617H}, or
pulsating delayed detonations \citep{1995ApJ...444..831H,1996ApJ...472L..81H} and could start
in the deep interior, from distributed hot spots, or near the surface.
The product is rich not only in radioactive $^{56}$Ni, but in intermediate-mass 
elements (such as Si, S, Ca, or Ar) and in non-radioactive iron-group isotopes.

Whatever the detailed mechanism and early (first $\sim$1 second) explosive history,
Type Ia supernovae have been studied for decades in the optical, ultraviolet, and near-infrared.
Various systematics (such as the ``Phillips" relation; Phillips 1993) have been uncovered 
and profitably used (Riess et al. 1998; Perlmutter et al. 1999), a range of peak 
brightnesses, atomic abundances, and $^{56}$Ni yields have been inferred, and
many detailed models of the explosion itself have been proffered.  However, though
a large fraction of the luminous energy of a Type Ia emerges in a hard X-ray continuum and 
gamma-ray lines, no Type Ia supernova has yet been observed and scrutinized at these energies.  While
the optical spectra and light curves require non-LTE (LTE: local-thermodynamic-equilibrium) and time-dependent radiative transfer 
of thousands of observed energy levels, depending collectively upon hundreds of thousands to millions of atomic
transitions for scores of ionization states (see, e.g., Pauldrach, Hoffmann, \& Hultzsch 2014), the gamma-ray signatures of Type Ia models
are relatively simple to generate and understand.  The energies and branching ratios of the finite
forest of gamma-ray lines produced in the $^{56}$Ni($\tau = 8.8$ days)$\rightarrow$$^{56}$Co($\tau = 111.3$ days)$\rightarrow$$^{56}$Fe 
decay sequence are well understood, as are the mean lifetimes ($\tau$) and the positron production factors.  The cross sections needed
for transfer calculations of the gamma-ray lines are simply those of (inelastic) 
Compton scattering, photoelectric absorption, and (below $\sim$70 keV) bremsstrahlung 
of the secondary electrons.  All models early in their expansion become homologous, 
freezing their density and abundance distributions. Therefore, the gamma-ray lines 
and hard-photon continuum fluxes as a function of time are ideal signatures of 
the geometry, abundances, abundance distributions, kinetic energies, and masses of the explosion.
In a very real sense, they ``X-ray" the debris, and provide the most direct fundamental 
constraints on Type Ia models. 


Recently, a Type Ia supernova, now designated SN 2014J, exploded in the galaxy Messier 82 at a distance
of $\sim$3.5 megaparsecs (Zheng et al. 2014; Goobar et al. 2014; Cao et al. 2014; Kuulers et al. 2014).  
Previous theory (e.g., Diehl \& Timmes 1998; Ambwani \& Sutherland 1988; Burrows 1990; Burrows \& The 1990; 
Burrows, Shankar, \& Van Riper 1991; Bussard, Burrows, \& The 1989; Chan \& Lingenfelter 1991;
H\"oflich et al. 1998a; Maeda et al. 2012; Milne et al. 2004; Summa et al. 2013) indicates that at such a close distance its gamma-ray
lines with energies of 812, 847, and 1238 keV might be detectable by INTEGRAL/SPI (Roques, Schanne, \& von Kienlin 2003; Vedrenne et al. 2003).
For a 10$^6$-second integration, INTEGRAL/SPI has an approximate narrow-line 3-$\sigma$ sensitivity at 1 MeV
of $3.1\times 10^{-5}$ photons cm$^{-2}$ s$^{-1}$. It also has a spectral resolution near 1.3 MeV of $\sim$2.5 keV, sufficient 
in principle to resolve the various gamma-ray lines whose widths have been estimated to be 10$-$20 keV 
(Burrows \& The 1990; Bussard, Burrows, \& The 1989).  Even more, the supernova's 
hard X-ray Compton continuum from $\sim$30 keV to $\sim$80 keV should easily be observed by 
NuSTAR, boasting a $10^6$-second sensitivity of a few $\times 10^{-8}$ photons cm$^{-2}$ s$^{-1}$ keV$^{-1}$ 
in this energy range (Harrison et al. 2013; Koglin et al. 2005).  The hard 
continuum of SN 2014J might also be observed by INTEGRAL/IBIS/ISGRI (Lebrun et al. 2003),
with a 3-$\sigma$ sensitivity from $\sim$20 to $\sim$100 keV of a few $\times 10^{-6}$ photons cm$^{-2}$ s$^{-1}$ keV$^{-1}$
\footnote{INTEGRAL/IBIS also has capabilities from $\sim$15 keV to $\sim$1 MeV, with the 
potential for detections at the $\sim$few $\times 10^{-5}$ cm$^{-2}$ s$^{-1}$ keV$^{-1}$ 
level between $\sim$100 and 1000 keV (Isern et al. 2013). The Suzaku HXD detector (Takahashi et al.2007) 
extends to $\sim$200 keV (and with reduced effective area to $\sim$600 keV) and might have a shot at detecting 
this supernova at early times (perhaps near $\sim$20 days after explosion) and at the higher continuum energies.}.
Excitingly, and by the end of 2014, ASTRO-H will be launched and will have capabilities in the region below 100 keV comparable to
those of NuSTAR.  Specfically, below $\sim$100 keV, HXI on ASTRO-H will have 
a 3-$\sigma$ sensitivity for a $10^6$-second integration of a few $\times 10^{-8}$ 
photons cm$^{-2}$ s$^{-1}$ keV$^{-1}$ (Kokubun et al. 2010). SGD on ASTRO-H has a design sensitivity 
over the 100$-$450 keV range of 5-10$\times 10^{-8}$ photons cm$^{-2}$ s$^{-1}$ keV$^{-1}$ for 
a $10^6$-second integration (Tajima et al. 2010).

There already exist in the literature many predictions for the hard photon signatures of many disparate Type Ia models (see above).
The gamma-ray line light curves, line shapes, hard X-ray continuum flux spectrum, flux band ratios, and the ratio
of the hard X-ray fluxes to those of the gamma-ray lines are together sensitive to and diagnostic of the $^{56}$Ni yields, $^{56}$Ni and high-Z element
spatial distributions and mixing, total masses, explosion kinetic energies, explosion asymmetries, and the distinction
between supersonic detonation and subsonic deflagration (for a review, see Burrows 1990). Depending upon model, 
the peak gamma-line flux occurs $\sim$60$-$90 days after explosion, the Comptonization continuum from $\sim$30 keV 
to $\sim$1 MeV peaks earlier (near day 20$-$30), and is usefully diagnostic of model before day 60, and the spectral 
peak of the continuum is found near $\sim$100$-$200 keV before day 100.  The fluxes below $\sim$100 keV are 
very senstive to photoelectric absorption in the debris, which in turn is related to iron-peak yield 
and its spatial distribution.  If there are iron-peak isotopes near the debris surface, the fluxes
below $\sim$100 keV can be severely photoelectrically suppressed, while the corresponding 
gamma-ray lines fluxes near $\sim$1 MeV are higher.  The Compton continuum flux above $\sim$100 keV 
would also be higher.  Hence, the ratios of the hard X-ray flux below $\sim$100 keV to the continuum 
flux above $\sim$100 keV or to the gamma line fluxes near $\sim$1 MeV are strongly 
dependent on the iron-peak distribution.  Correspondingly, if the $^{56}$Ni is 
more deeply buried in the debris, the hard X-ray fluxes below $\sim$100 keV are enhanced 
and these systematics are reversed.  Moreover, if the explosion energies are large, and/or 
the $^{56}$Ni yields high, the $^{56}$Co-decay line fluxes at 847 and 1238 keV can be 50$-$100\% higher.
The line shapes directly reflect the distribution of the produced $^{56}$Ni in velocity space 
(Burrows \& The 1990).  In fact, an analytic model for these line shapes, including Doppler shift effects and distinguishing
the matter and $^{56}$Ni distributions, has been developed (Bussard, Burrows, \& The 1989).

We surmise, however, that a focussed set of new calculations in aid of the ongoing SN 2014J campaign to detect
its hard continuum and line emissions might be of use.  In this spirit, we calculate in this paper various theoretical
hard-photon signatures for a representative set of Type Ia explosion models, put at the 3.5 megaparsec distance of M82.  
The results can be scaled to any other distance using the inverse square law.
The model results are also provided in tabular form (at \url{http://www.astro.princeton.edu/\~{}burrows/sn2014j/})
to help observers engaged in measurement and interpretation efforts navigate the theoretical terrain.
In this paper, we don't provide gamma-ray line profiles, but encourage the interested reader to 
explore the associated predictions of Burrows \& The (1990) and the analytic model of Bussard, Burrows, \& The (1989).

\section{Method}
\label{method}

To derive the emergent spectrum at a given post-explosion time we employ the 
variance reduction Monte Carlo code developed in Burrows \& The (1990) and The, Burrows, \& Bussard (1990),
augmented to include the bremsstrahlung X-ray production of the Compton electrons 
and the iron fluorescence line at $\sim$6.4 keV (Clayton \& The 1991; The, Bridgman, \& Clayton 1994). The six 
most prominent lines of $^{56}$Ni decay (including the 158, 750, and 812 keV lines) and the 
forty-five most prominent lines of $^{56}$Co decay (the most important of which are at 847, 1238, 1772, and 2598 keV)
are followed from emission, through Compton scatterings, to either photoelectric absorption
in the homologously-expanding debris or escape\footnote{We also include the ten lines of 
$^{57}$Co decay, for which the 14.4, 122, and 136 keV lines are the most important.  For model W7,
the mass of $^{57}$Ni is $4.81\times 10^{-2}$ M$_{\odot}$.  For models W7E and W7A, 
the mass of $^{57}$Ni is $8.5\times 10^{-3}$ M$_{\odot}$.  For model w7dn, 
the mass of $^{57}$Ni is $2.60\times 10^{-2}$ M$_{\odot}$ and for model w7dt
it is $2.72\times 10^{-2}$ M$_{\odot}$.  For all other models, 
we assume a solar ratio for $^{57}$Ni/$^{56}$Ni of $\frac{1}{41}$. At these levels, 
the continuum results before day 100 are very little affected.}.  The production of positrons
is included, but they are assumed to form positronium instantaneously and 
to annihilate in situ (Bussard, Ramaty, \& Drachman 1979).  Both two-photon 
(25\%) and three-photon (75\%) decays are allowed, and those photons are then followed
in the Monte Carlo.  The electron-positron pair-production cross section is taken 
from Ambwani \& Sutherland (1988), the full angle-dependent Klein-Nishina formula for Compton scattering is used,
and the photoelectric cross sections for the thirty-one most abundant elements thought to reside
in Type Ia debris are taken from Veigele et al.(1973) and Henke et al. (1982).
Since the matter speeds are small compared with the speed of light, we don't 
include Doppler shifts due to matter motions (thermal or bulk) in the calculations.  
These would need to be included if we were to focus on the detailed line profiles, 
but for the integrated line emissions they are not germane.

The bremsstrahlung calculation proceeds as follows: For each gamma-ray photon 
from the decays of $^{56}$Ni and $^{56}$Co that Compton scatters, its new direction and 
energy are calculated by Monte Carlo sampling using the Klein-Nishina 
differential cross section. At the same time, the recoil kinetic energy 
of the electron and its location are recorded for post-processing. 
When post-processing to derive the bremsstrahlung flux, we assume 
the energetic electrons slow down by inelastic collisions with atoms and plasma ions 
where they were Comptonized.  For each recorded Comptonized electron, 
we calculate the bremsstrahlung spectrum generated in the medium (see Clayton \& The 1991).
Then, we follow the propagation of the bremsstrahlung photons using the same
Monte Carlo procedure used for the gamma-ray lines.  Photoelectric absorption, electron collisions,
and electron-capture on $^{56}$Fe parent's nuclei create shell vacancies in ions.  Each K-shell vacancy 
is followed by K X-ray fluorescence and the number of K X-ray line photons 
generated is calculated by multiplying the number of K-shell vacancies 
by the K-shell fluorescence yield of the absorbing atoms (Bambyneck et al. 1972).  
The number of K X-ray line photons that emerges is then calculated by
the self-same Monte Carlo method. The, Bridgman, \& Clayton (1994) estimate that 
most of the 6.4-keV Fe K-shell X-ray production is due in fact to the 
electron-capture process, so that since we do not include ionization 
due to electron collisions our calculated Fe K X-ray line fluxes 
are probably accurate to $\sim$10\%.

All the explosion models we study are spherical, though we have observed that the differences 
in gamma-ray signatures between aspherical (e.g., Summa et al. 2013; Maeda et al. 2012) 
and spherical models are not as large as the differences between the vast array of published spherical models, with their
disparate $^{56}$Ni yields and distributions, heavy-element distributions, kinetic energies, 
and burning regimes.  We use the element and density distributions provided by the original 
Type Ia model builders (except in the few cases when we have artifically mixed the ejecta),
and have endeavored in choosing the subset of models we highlight here to cover as wide a range
of anticipated behaviors as possible.  

Therefore, for this SN 2014J study we have chosen fifteen models in the literature 
to represent the range of Type Ia explosion models.  They span the model space
discussed over the last 30 years, though are all spherical realizations,
and include deflagrations (W7; \citet{w7}, along with mixed variants W7E and W7A), 
delayed detonations (dd4 $-$ \citet{WW91}; w7dn, w7dt $-$ \citet{1992ApJ...393L..55Y}; 
m36 $-$ \citet{1995ApJ...443...89H}; dd202c $-$ \citet{1998ApJ...495..617H}), 
pulsating delayed detonations (pdd54 $-$ \citet{1995ApJ...444..831H}), 
merger detonations (det2, det2e2, det2e6 $-$ \citet{1993A&A...270..223K,1996ApJ...472L..81H}), 
and sub-Chandrasekhar helium detonations (hed6, hed8, hecd $-$ \citet{1996ApJ...472L..81H}).  

\begin{figure}
\begin{center}
\includegraphics[height=.55\textheight,angle=90]{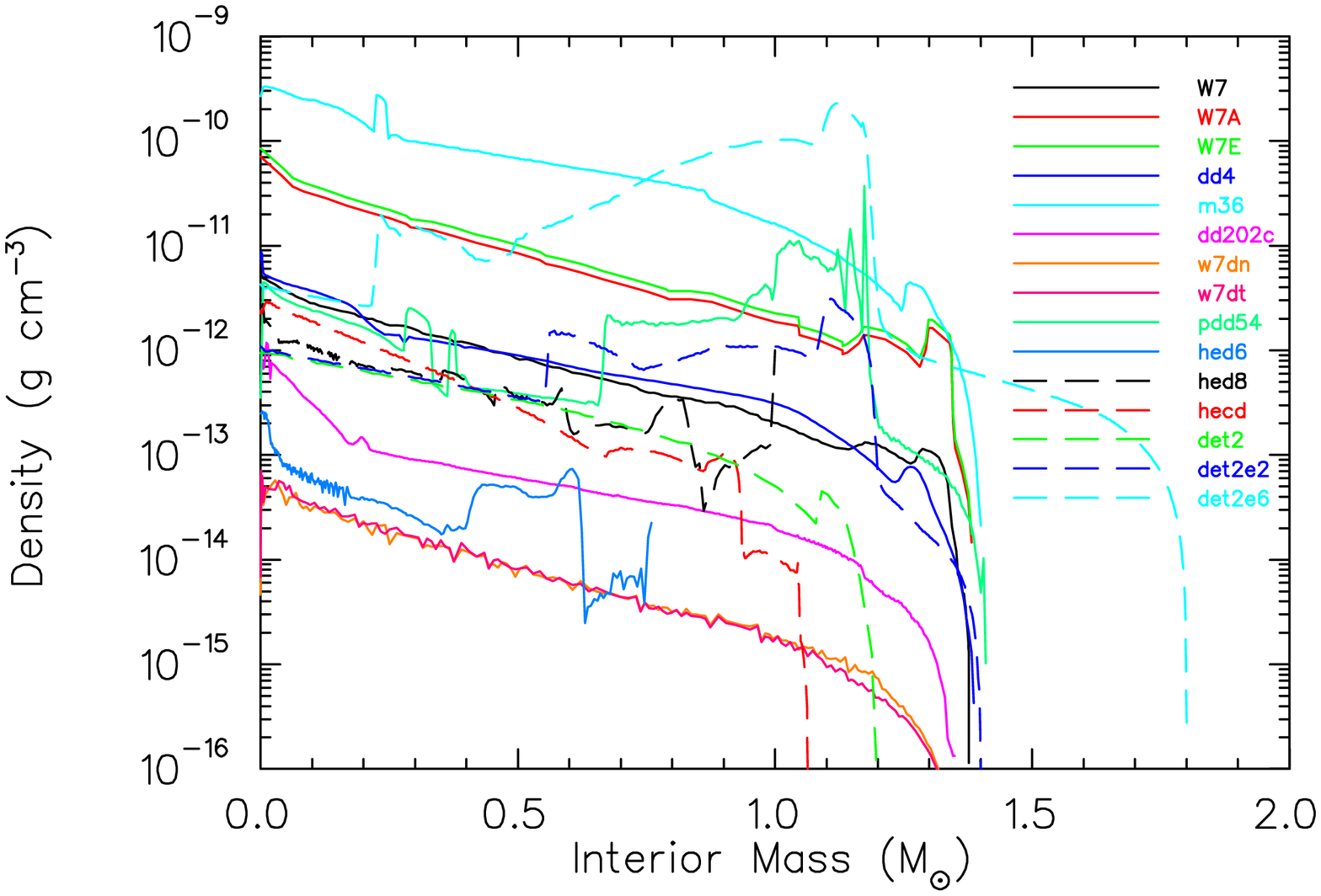}
\end{center}
\caption{The density (in g cm$^{-3}$) versus interior mass (in M$_{\odot}$)
at day 10 for the fifteen models studied in this paper. For clarity of presenation,  W7A is multiplied by 10,
W7E is multiplied by 12, m36 is multiplied by 100, w7dn and w7dt are divided by 100, and hed6 is divided by 10. 
Note that the expansion is homologous and that the density at a given interior mass scales 
as $\frac{1}{t^3}$, where $t$ is the time since the start of explosion.  
}
\label{fig0}
\end{figure}

\begin{figure}
\begin{center}
 \includegraphics[height=.55\textheight,angle=90]{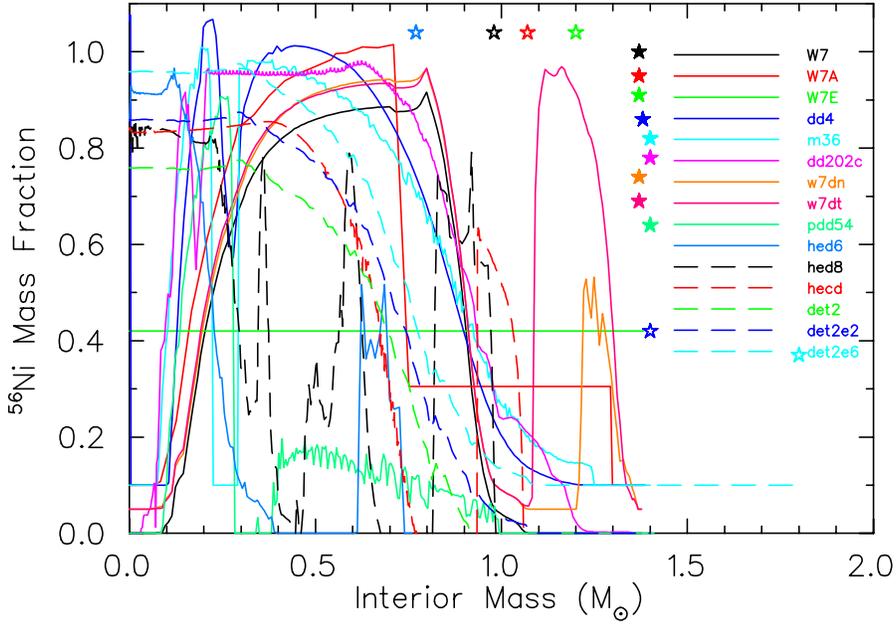}
\end{center}
\caption{The $^{56}$Ni fraction versus interior mass (in M$_{\odot}$) for the fifteen
models highlighted in this paper.  This figure depicts the $^{56}$Ni distribution for each model.  
To attempt a degree of clarity in what would otherwise be even more of a jumble, we have
shifted the fractions for W7A by -0.1, dd4 by +0.1, m36 by +0.1, w7dn by +0.05,
w7dt by +0.05, hed8 by -0.1, hecd by -0.1, det2 by -0.1, and det2e6 by + 0.1.
Note that the various models have different total ejecta masses, a fact reflected 
by the different positions of the star symbols with the associated model colors that identify
the outer boundary mass of a given model.  See Table \ref{specs}.
}
\label{fig0_dist}
\end{figure}

W7 is a fiducial Chandrasekhar deflagration model with an explosion energy of $1.3\times 10^{51}$ ergs
and a $^{56}$Ni mass of 0.58 M$_{\odot}$ that seems to fit well the spectra and light curves 
of normal Type Ias.  Model W7E fully mixes the $^{56}$Ni, and model W7A partially mixes it. dd4 
is a delayed detonation with an explosion energy of $1.24\times 10^{51}$ ergs, a 
total mass of 1.39  M$_{\odot}$, and a $^{56}$Ni mass of 0.61 M$_{\odot}$.  w7dn is a 
delayed detonation model of 1.37 M$_{\odot}$ with an explosion energy of $1.6\times 10^{51}$ ergs, 
constructed to have the same core distribution of $^{56}$Ni as W7, but given an extra $^{56}$Ni surface 
cap of 0.04 M$_{\odot}$. Hence, model w7dn has a total $^{56}$Ni mass of 0.62 M$_{\odot}$.
w7dt is the same as w7dn, but with an explosion energy of $1.61\times 10^{51}$ ergs and 
an extra outer layer of 0.18 M$_{\odot}$ of $^{56}$Ni, for a total $^{56}$Ni mass of 0.76 M$_{\odot}$.
m36 is a delayed detonation model with an explosion energy of $1.51\times 10^{51}$ ergs 
and 0.59 M$_{\odot}$ of $^{56}$Ni. dd202c is a delayed detonation model with 
an explosion energy of $1.27\times 10^{51}$ ergs and a continuous $^{56}$Ni
distribution from the core to an interior mass point of 1.2 M$_{\odot}$,
with a total mass of 1.40 M$_{\odot}$ and a $^{56}$Ni mass of 0.72 M$_{\odot}$.
pdd54 is pulsating delayed detonation model with an explosion energy of $1.02\times 10^{51}$ ergs, 
in which slow deflagration leads to burning inefficient enough to induce pulsation
before detonation. This model has a total $^{56}$Ni mass of 0.17 M$_{\odot}$.
det2 is a merger detonation model having an explosion energy of $1.54\times 10^{51}$ ergs, 
a total mass of 1.20 M$_{\odot}$, and a $^{56}$Ni mass of 0.62 M$_{\odot}$. det2e2 is a 
merger detonation model with an explosion energy of $1.44\times 10^{51}$ ergs, a total 
mass of 1.40 M$_{\odot}$, and a $^{56}$Ni mass of 0.63 M$_{\odot}$.  det2e6 is a merger 
detonation model with an explosion energy of $1.43\times 10^{51}$ ergs, a total mass 
of 1.80 M$_{\odot}$, and a $^{56}$Ni mass of 0.63 M$_{\odot}$. hed6 is a 
He-detonation model with an explosion energy of $0.72\times 10^{51}$ ergs, a total 
mass of 0.77 M$_{\odot}$, and a $^{56}$Ni mass of 0.26 M$_{\odot}$. hed8 is a 
He-detonation model with an explosion energy of $1.03\times 10^{51}$ ergs, a total mass 
of 0.96 M$_{\odot}$, and a $^{56}$Ni mass of 0.51 M$_{\odot}$. Finally, hecd is a superluminous helium-detonation 
model with an explosion energy of $1.3\times 10^{51}$ ergs, a total mass of 1.07 M$_{\odot}$, 
and a $^{56}$Ni mass of 0.72 M$_{\odot}$.  These last three sub-Chandrasekhar detonation models 
(hed6, hed8, and hecd) all have central $^{56}$Ni concentrations, as well as outer surface 
layers of $^{56}$Ni. All together, these fifteen models span the model space rather well.
We plot the fifteen model density profiles in Figure \ref{fig0} (at day 10) and 
their $^{56}$Ni distributions in Figure \ref{fig0_dist}. Figure \ref{fig0} demonstrates
the wide range in density profiles represented, from the high densities of model det2e6 to the much lower densities
of model hed6.  Note the enhancements in the outer densities of models det2e2 and pdd54.  Table \ref{specs} summarizes
the basic model specifications.

\begin{table}[tbhp]
\caption{Type Ia Explosion Model Characteristics}
\begin{center}
\begin{tabular}{l|l|l|l|l}
\hline
Model   &  Total Mass   & $^{56}$Ni Mass & Explosion Energy & Reference\\
      & (M$_{\odot}$)) &(M$_{\odot}$)        & (10$^{51}$ ergs) & \\
\hline
W7   & 1.38 & 0.58 & 1.3 &  1\\
W7E  & 1.38 & 0.58 & 1.3 &  1\\
W7A  & 1.38 & 0.58 & 1.3 &  1\\
dd4 & 1.39 & 0.61 & 1.24 &  2\\
w7dn & 1.37 & 0.62 & 1.6 &  3\\
w7dt & 1.37 & 0.76 & 1.61 & 3 \\
m36 & 1.39 & 0.59 & 1.51 &  4\\
dd202c & 1.4 & 0.72 & 1.27 & 5 \\
pdd54& 1.38 & 0.17 & 1.02 &  6\\
det2 & 1.2 & 0.62 & 1.54 &  7\\
det2e2 & 1.4 & 0.63 & 1.44 &  7\\
det2e6& 1.8 &0.63 & 1.43 &  7\\
hed6& 0.77 & 0.26 & 0.72 &  7\\
hed8 & 0.96 & 0.51 & 1.03 &  7\\
hecd & 1.07 & 0.72 & 1.3 &  7\\
\hline
\end{tabular}
\end{center}
\tablerefs{
(1) Nomoto, Thielemann, \& Yokio (1984);
(2) Woosley \& Weaver(1991);
(3) Yamaoka et al. (1992);
(4) H\"oflich (1995);
(5) H\"oflich et al.(1998b);
(6) H\"oflich et al. (1995);
(7) H\"oflich et al. (1996).
}

\label{specs}
\end{table}

A few approximate facts are worth noting.  The peak gamma line fluxes occur near a time, $t_p \propto$ ($\kappa_{\gamma} \frac{M^2\tau}{E})^{1/3}$,
with a peak flux $F_p \propto \frac{1}{\tau} e^{-\frac{3t_p}{2\tau}}$, where $M$ is the debris mass, $E$ is the explosion kinetic energy,
$\tau$ is the mean life of the decay, and $\kappa_{\gamma}$ is the Compton opacity of the line.  
Bussard, Burrows, \& The (1989) have demonstrated that this behavior generally tracks the calculated numerical behavior rather well. The
approximate number of Compton scatterings of an $\sim$MeV gamma ray necessary to downscatter to
a photon energy of $\varepsilon_{\gamma}$ is $\sim$$\frac{m_ec^2}{\varepsilon_{\gamma}}$.  So, to achieve $\varepsilon_{\gamma} = 50$ keV
requires $\sim$10 scatterings.  The Compton cross sections at 847, 1238, and 3200 keV are 
0.344$\sigma_T$, 0.285$\sigma_T$, and 0.166$\sigma_T$, respectively, where $\sigma_T$ is the Thomson cross section. 
After $\sim$150 days, much of the continuum flux below $m_ec^2 = 511$ keV is due to the three-photon decay of positronium,
and below the 122 and 136 keV lines of $^{57}$Co due to their Comptonization (if $^{57}$Ni is present).

\section{Results}
\label{results}

\begin{figure}
\begin{center}
\includegraphics[height=.55\textheight,angle=90]{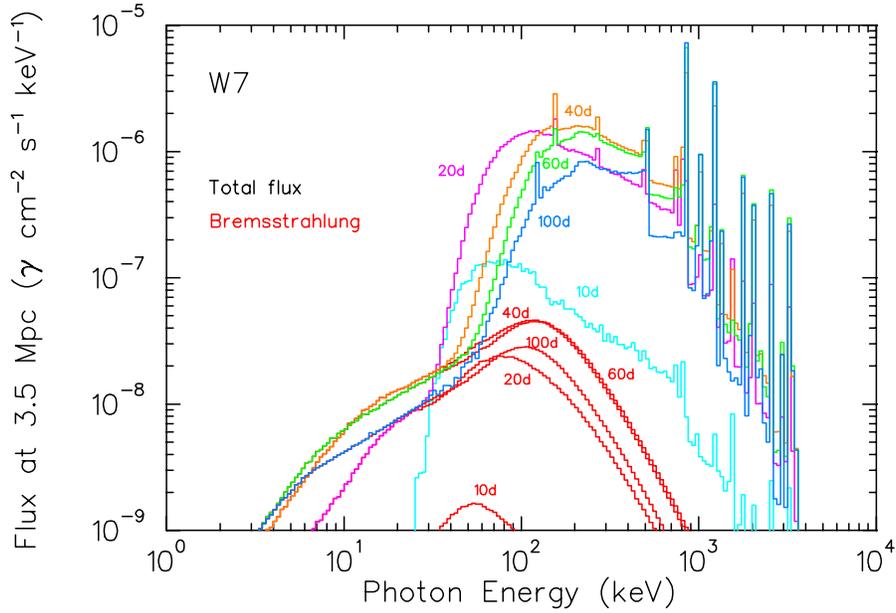}
\end{center}
\caption{The flux spectra at a distance of 3.5 Mpc in photons cm$^{-2}$ s$^{-1}$ keV$^{-1}$ for 
the emergent hard X-ray photons and gamma-ray lines 10, 20, 40, 60, and 100 days
after explosion for the fiducial model W7.  The photon energy
is in keV from 1 keV to $\sim$3.2 MeV. Included are curves for the total flux and 
its bremsstrahlung contribution, indicating the importance of bremsstrahlung below $\sim$60 keV.
Note that the line fluxes depicted are per plotting bin width, and are not resolved.  They show the line positions,
but the relative heights are only crude measures of the actual fluxes, whose integrals (the total line fluxes) are shown
correctly in photons cm$^{-2}$ s$^{-1}$ in Figures \ref{fig3}, \ref{fig4}, and \ref{fig5}.
}
\label{fig1}
\end{figure}

\begin{figure}
\begin{center}
\includegraphics[height=.55\textheight,angle=90]{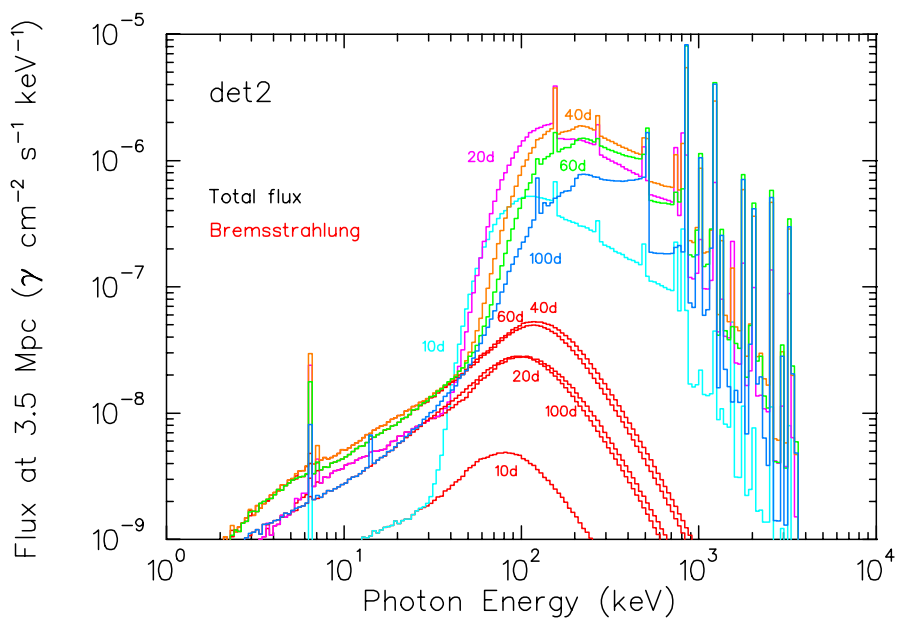}
\end{center}
\caption{The same as Figure \ref{fig1}, but for model det2. Notice that the continuum
and line fluxes for this detonation model at the early phases (e.g., 10 days) are much 
higher at higher photon energies, and lower at lower photon energies.  The former is due to the more rapid 
expansion, that liberates the hardest photons more quickly, while the latter is due to greater
photoelectric absorption by the enhanced $^{56}$Ni mass and the presence of iron-peak elements near the periphery.
The presence of iron-peak nuclei in the outer regions results in a visible iron 
fluorescence line at $\sim$6.4 keV, absorbed in W7 (Figure \ref{fig1}).}
\label{fig1_det2}
\end{figure}

\begin{figure}
\begin{center}
\includegraphics[height=.55\textheight,angle=90]{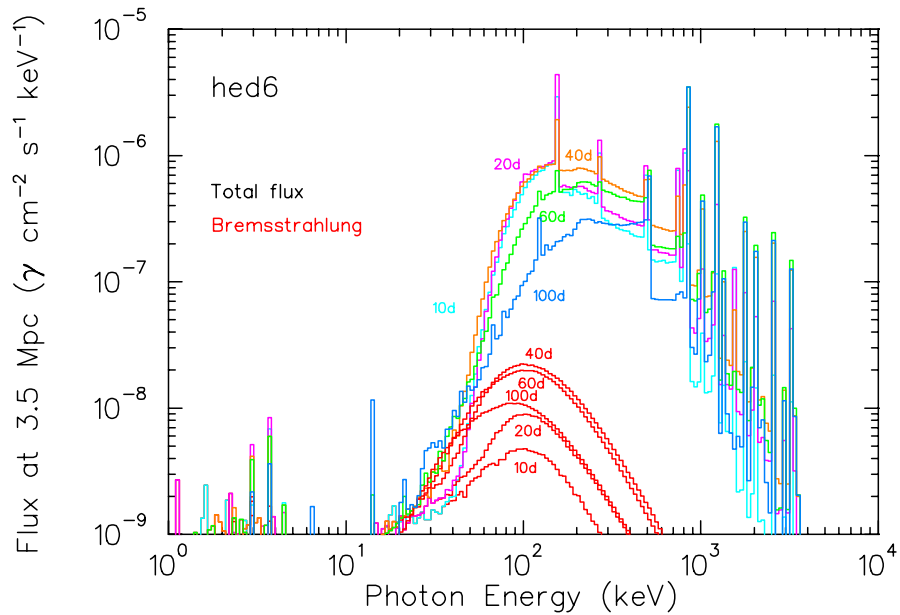}
\end{center}
\caption{The same as Figure \ref{fig1}, but for detonation model hed6.
Note that the band fluxes below $\sim$50 keV are even more suppressed than those for model det2 (Figure \ref{fig1_det2}).
This is due to photoelectric absorption by the $^{56}$Ni cap at the very periphery of this model.
Note also that concommitantly for this model the hard X-ray continuum above $\sim$60 keV and the gamma-line fluxes
are significantly higher at day 10, but lower at day 100.  The former reflects, among other things, 
the lower total mass of hed6, while the latter reflects the higher $^{56}$Ni yield of model det2.
}
\label{fig1_hed6}
\end{figure}

\begin{figure}
\begin{center}
\includegraphics[height=.55\textheight,angle=90]{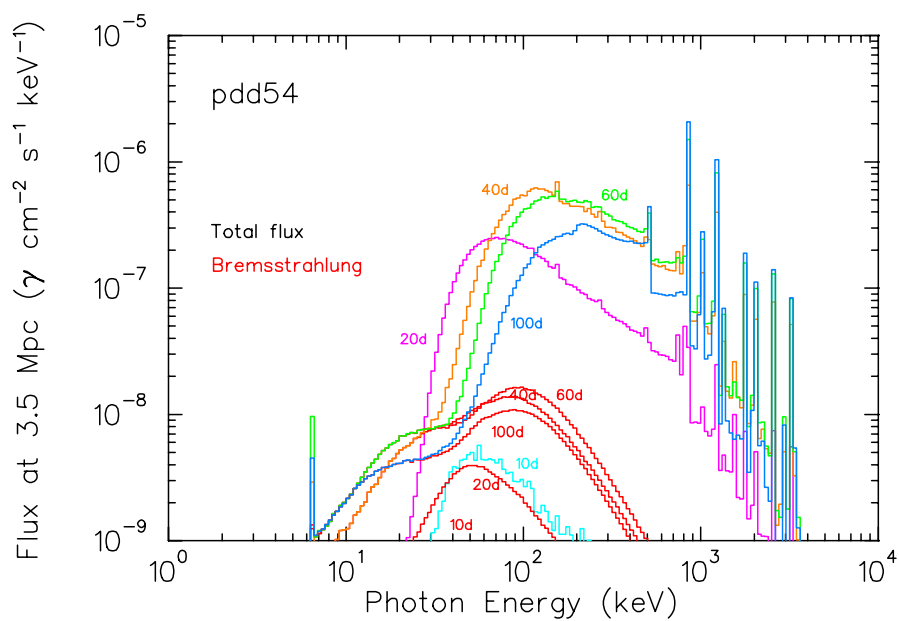}
\end{center}
\caption{The same as Figure \ref{fig1}, but for pulsating delayed detonation model pdd54.
The low flux values are reflective mostly of the low $^{56}$Ni burden in this model.
Note that the flux for this model at day 20 is approximately three times lower
at $\sim$70 keV than the corresponding flux for model W7, and that the photon energy
at peak has also shifted to lower energies.
}
\label{fig1_pdd}
\end{figure}

Rather than plot all the results for all fifteen models, we highlight a few representative models
with which to demonstrate various common effects.
Figures \ref{fig1}, \ref{fig1_det2}, \ref{fig1_hed6}, and \ref{fig1_pdd} depict the flux spectra of
the emergent hard X-ray photons and gamma-ray lines at various epochs after explosion from 10 to 100 days
for representative deflagration (W7), merger detonation (det2), helium detonation (hed6),
and pulsating delayed detonation (pdd54) models.  
Note that the continuum fluxes for many models peak near $\sim$200 keV 
and day 40 (except for the pdd54 model), but that the deviations from this behavior are discriminating.  
The calculations include the bremsstrahlung contributions of the Compton secondaries. 
For all models, the continuum flux peak slides to higher energies with time, while the spectral slope 
between $\sim$40 and $\sim$100 keV (always positive) decreases.  The (mostly) bremsstrahlung fluxes 
below $\sim$30 keV have positive slope near $\sim$1.0.
As these figures suggest, if the iron-peak elements do not reside in the outer zones of the debris 
in abundance (as they do, for instance, for models hed6, w7dn, and w7dt), bremsstrahlung below $\sim$60 keV can be quite important.  
This will be particularly relevant to the interpretation of anticipated NuSTAR data.   The continuum and line fluxes for merger 
detonation model det2 are higher earlier than for model W7, particularly earlier than day 10, but are similar 
to those for model W7 at later times, while W7's peak fluxes for the hard X-rays below 100 keV 
are higher than those for model det2.

For model hed6, the fluxes below $\sim$40 keV are suppressed with respect to those for models W7 and det2.  
This is due to photoelectric absorption by the $^{56}$Ni cap at the periphery of model hed6 and to its smaller ejecta mass. 
Interestingly, at day 10 and earlier, the fluxes at energies above 50 keV for model hed6 are higher than those for model W7,
while later than day 60 its line and continuum fluxes are lower than those for model W7 (reflecting hed6's lower $^{56}$Ni mass).
The generally low flux values for the pulsating delayed detonation model pdd54 shown in Figure \ref{fig1_pdd}
reflect the low $^{56}$Ni yield in this model. Note that the flux at $\sim$70 keV for this model at 
day 20 is approximately three times lower than the corresponding flux for model W7, but that 
near $\sim$30 keV the fluxes for these two models are comparable.  The slow expansion speed of model pdd54 and high cap densities 
leave the densities higher longer and scatters more photons at these times to lower 
energies, but the lower $^{56}$Ni yield partially compensates.  The photon energy at peak 
near day 40 has also shifted from the $\sim$200 keV for many models to $\sim$100 keV 
for model pdd54, and this model has $\sim$20\% to a factor of $\sim$3$\times$ lower flux.

\begin{figure}
\begin{center}
\includegraphics[height=.55\textheight,angle=90]{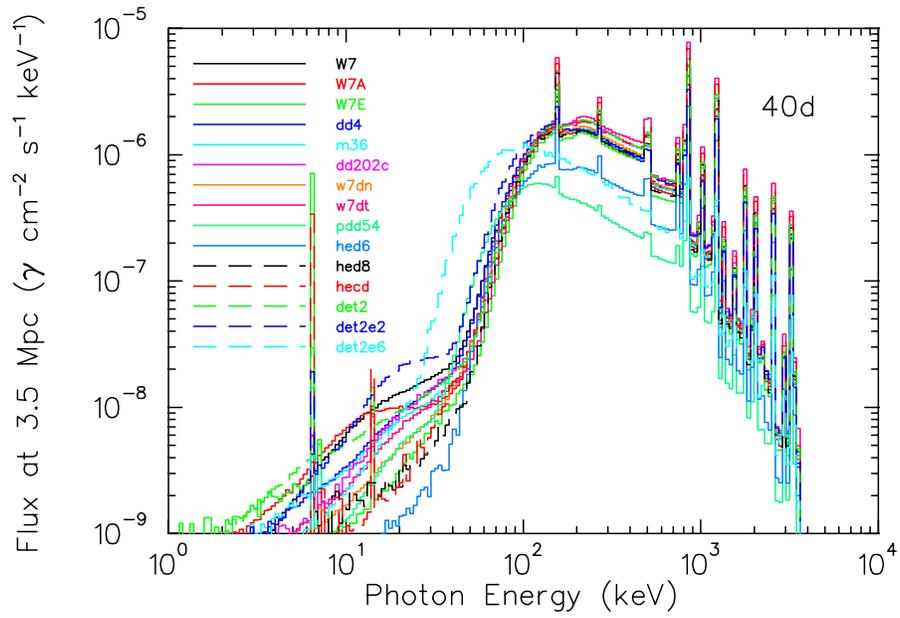}
\end{center}
\caption{The same as Figure \ref{fig1}, but for all fifteen models at day 40.
The model-to-model variation of fluxes at this epoch (as well as at others) is quite wide throughout 
the spectral interval depicted, but is particularly interesting in the NuSTAR bands.
For completeness, we also include the 6.4-keV fluorescence line of iron.
}
\label{fig1_all_40}
\end{figure}

\begin{figure}
\begin{center}
\includegraphics[height=.55\textheight,angle=90]{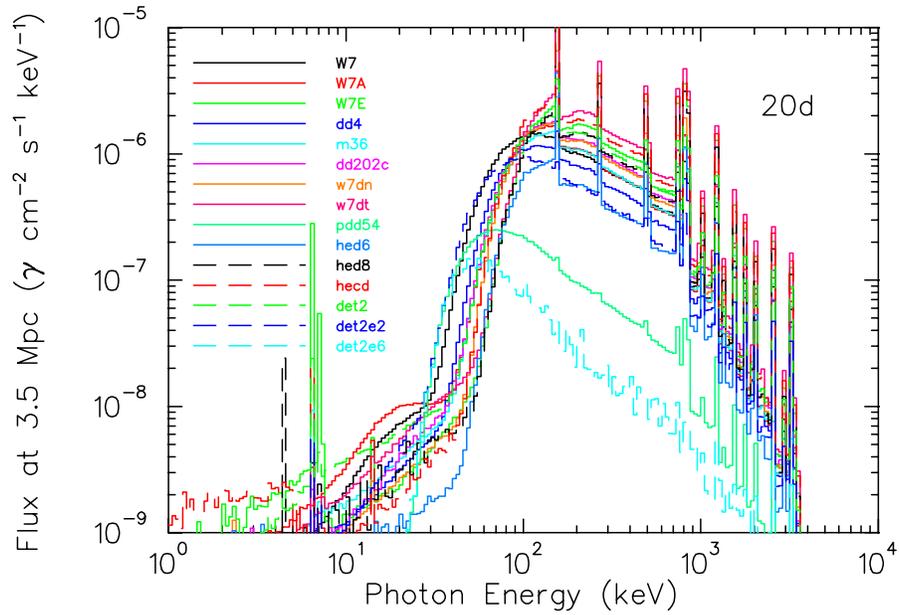}
\end{center}
\caption{The same as Figure \ref{fig1_all_40}, but for all fifteen models at day 20.
This figure clearly demonstrates the diagnostic differences between the various models at all 
energies.
}
\label{fig1_all_20}
\end{figure}

Figures \ref{fig1_all_40} and \ref{fig1_all_20} compare the spectra at days 40 and 20 
for all fifteen models investigated, summarizing some of the clear spectral differences between them. We call attention to the 
high fluxes for model det2e6 below $\sim$100 keV at day 40, caused by its relatively dense ejecta
(an efficient converter of MeV lines to X-ray continuum by Compton scattering), but also the low fluxes
above $\sim$40 keV for this same model at day 20.  As Figure \ref{fig1_all_40} demonstrates, 
at day 40 the continuum fluxes below $\sim$40 keV and above $\sim$200 keV
and the line fluxes in the MeV range distinguish the various models most clearly.
However, as Figure \ref{fig1_all_20} demonstrates, the continuum and line fluxes 
in all energy ranges at earlier times (such as day 20) are even more discriminating.

\begin{figure}
\begin{center}
\includegraphics[height=.55\textheight,angle=90]{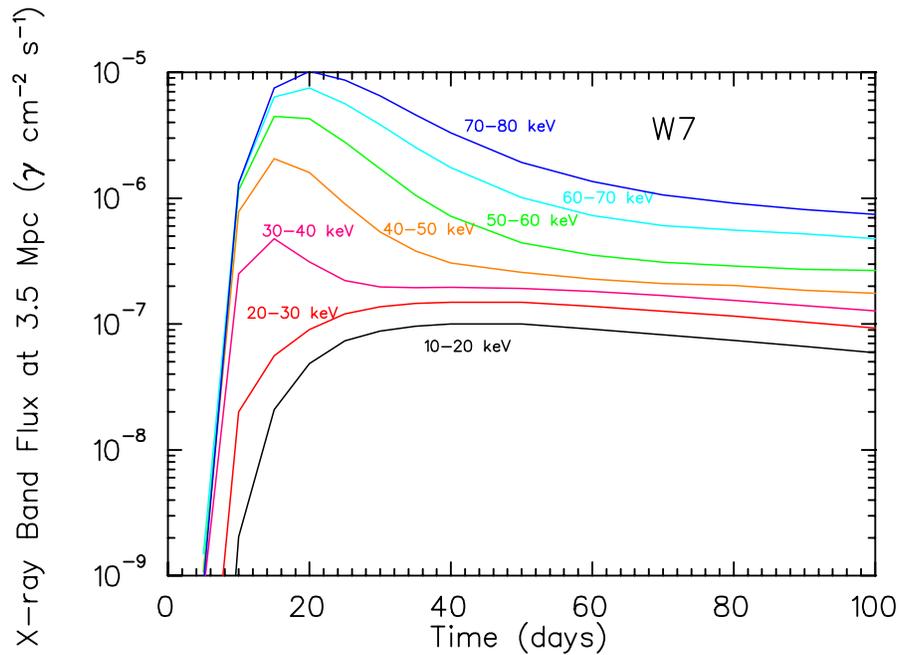}
\end{center}
\caption{The light curves of hard X-rays in 10-keV-width bins for the representative deflagration model W7.
The curves include bremsstrahung by Compton electrons.  Note that the curves peak early during 
the Type Ia supernova development, near day 20, that the hardest bands are generally the brightest (below $\sim$100 keV),
and that the curves decay rather slowly after peak.
} 
\label{fig2}
\end{figure}

\begin{figure}
\begin{center}
\includegraphics[height=.55\textheight,angle=90]{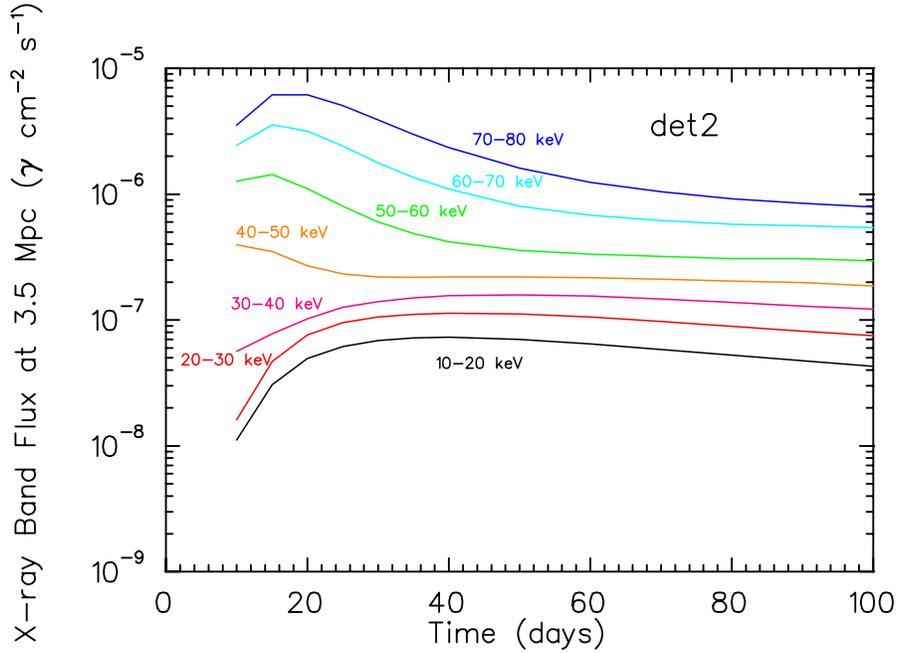}
\end{center}
\caption{The same as Figure \ref{fig2}, but for detonation model det2.
Note that these band fluxes are suppressed at later times relative to those for model W7 (Figure \ref{fig2})
by the enhanced photoelectric absorption in model det2. However, due to the rapid expansion,
all the band fluxes for det2 are higher than in fiducial model W7 at the earliest epochs after explosion.  This is one 
diagnostic signature of the rapid disassembly of detonations, as opposed to deflagrations, and the proximity
of $^{56}$Ni to the periphery in the former.
} 
\label{fig2_det2}
\end{figure}

\begin{figure}
\begin{center}
\includegraphics[height=.55\textheight,angle=90]{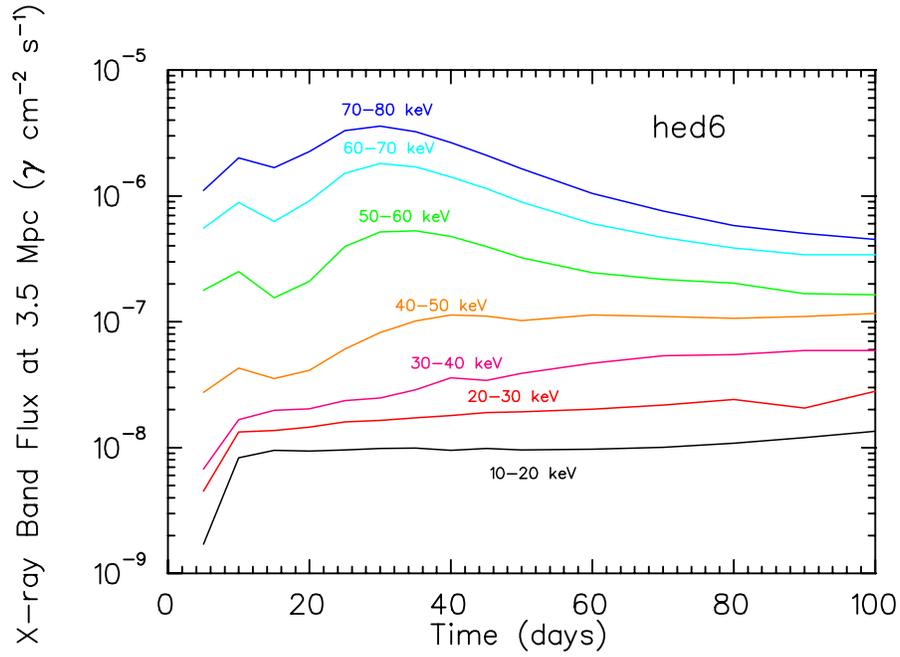}
\end{center}
\caption{The same as Figure \ref{fig2}, but for detonation model hed6.
}
\label{fig2_hed6}
\end{figure}

\begin{figure}
\begin{center}
\includegraphics[height=.55\textheight,angle=90]{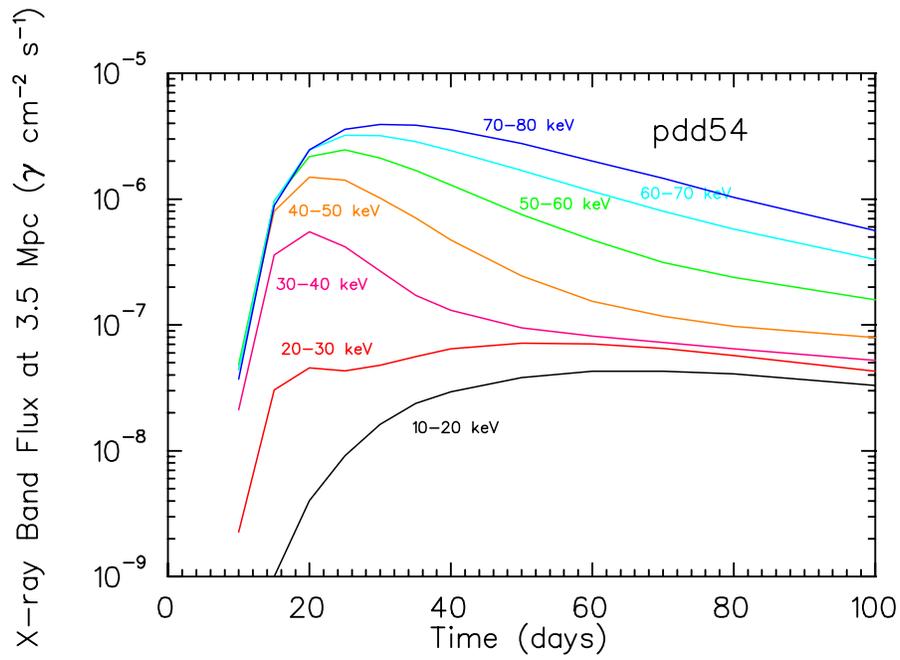}
\end{center}
\caption{The same as Figure \ref{fig2}, but for pulsating delayed detonation model pdd54.
}
\label{fig2_pdd}
\end{figure}

Figures \ref{fig2}, \ref{fig2_det2}, \ref{fig2_hed6}, \ref{fig2_pdd} portray the light curves of hard X-rays 
in 10-keV-width bins for the representative models W7, det2, hed6, and pdd54. These plots are particularly 
relevant to NuSTAR.  Bremsstrahlung is important below $\sim$60 keV, particularly at later times.
Note that the curves peak (depending upon the energy bin) early during a Type Ia supernova. 
For model W7, they peak near day 20; for model det2 they peak between days 10 to 30 (depending upon energy bin); 
for model hed6, there is an early, weaker peak, followed by a brighter peak from 50 to 80 keV near day 30; 
and model pdd54 peaks between days 20 and 30. The hardest bands below $\sim$100 keV are generally the brightest, 
and the curves decay rather slowly after peak. In fact, these band fluxes are roughly constant from day 60 to after day 100 
for models W7, det2, and hed6 $-$ less so for model pdd54.  For model det2, due to its rapid 
expansion and more outward distribution of $^{56}$Ni, the band fluxes for det2 peak earlier than in model W7.
This is one diagnostic of the rapid disassembly of detonations, as opposed to deflagrations, and the proximity
of $^{56}$Ni to the periphery in the former.  However, above 30 keV and at later times the band fluxes 
of models det2 and W7 are comparable.  As Figure \ref{fig2_hed6} indicates (and as noted earlier), the 
hard X-ray band fluxes for helium detonation model hed6 are much higher than those for
model W7 before day 10, though thereafter those for model W7 quickly exceed those for model hed6.  
Note that the early fluxes for model pdd54 are quite different from those for model hed6, 
but that near day 40 they are comparable in the 70$-$80 keV band. These examples (among others) 
emphasize that a time series (in addition to a spectrum) is important to properly discriminate 
between the very different models.  Though not accessible to NuSTAR, in anticipation of the launch of ASTRO-H, with its
SGD detector with sensitivity in the 200 to 450 keV range, we provide in Figure 13
light curves in this broad band for all fifteen models.  As noted earlier, the fluxes
in this region of the spectrum can be quite diagnostic of model.

Figures \ref{fig3} and \ref{fig4} plot the light curves (in days after explosion)
of the 847 and 1238 keV lines of $^{56}$Co decay at a distance of 3.5 Mpc
for the entire suite of fifteen models with which we have chosen to depict the wide range 
of Type Ia explosion models in the literature. These curves are most germane to INTEGRAL 
and its SN 2014J campaign. The peak emission occurs for most models approximately between day 60  
and day 100, but the early rise of a light curve is a stiff function of model. 
For instance, at day 40 these line fluxes vary from model to model by an order of magnitude, 
and the fluxes at day 20 vary by more than two orders of magnitude. At late times, the 847 and 1238 keV line fluxes 
reflect the bulk $^{56}$Ni yield.  In this sense, some our models overlap the line 
predictions of Maeda et al. (2012). Note that merger detonation model det2e6, with its larger 
total mass (1.80 M$_{\odot}$) and ``anomalous" mass distribution (Figure \ref{fig0})
has, as a result, quite muted line fluxes during the first hundred days, but larger Compton continnum
fluxes below $\sim$100 keV.

Figure \ref{fig5} depicts the corresponding light curve for the 812 keV line of $^{56}$Ni.
The peak fluxes differ from model to model by approximately a factor of one hundred and occur between days 10 and 30. 
This is much earlier than the corresponding peaks for the  847 and 1238 lines of $^{56}$Co 
(as well as for its other lines), and the 812 keV line flux
decays much faster for all models.  The early emergence and fast decay of the 812 keV line 
are mostly due to the shorter decay time of $^{56}$Ni ($\tau = 8.8$ days), while the lower
fluxes reflect the greater Compton opacities at these earlier (denser) times.  Therefore, the strength of
this line is an important signature of the rapidity with which the debris expands and becomes transparent
to Compton scattering.  As a result, it is an important diagnostic of the expansion
speed and the total mass of the ejecta, as well as of the presence of $^{56}$Ni in the outer zones, and can 
discriminate well between most detonation and deflagration models.  However, the lower
associated fluxes make it more difficult to capture.

Figure \ref{fig6} combines aspects of both the continuum and line signatures of the various
models to provide a good metric with which to distinguish them.  It depicts the evolution of the 
ratio of the 847 keV line flux to the total flux in a 40 to 80 keV bin for
all fifteen models studied in this paper.  Such line$-$continuum flux ratios (others can easily be envisioned)
are very diagnostic of model, varying as they do from model to model by factors of more than 
one hundred, depending upon epoch.  A number of models have
high continuum fluxes below $\sim$100 keV when they also have low line fluxes near $\sim$1 MeV 
(as well as low continuum fluxes from $\sim$100 to 1000 keV), 
and vice versa.  So, such ratios amplify the model differences rather well.  Their 
usefulness does, however, depend upon getting good data for both line and continuum. 
Figure \ref{fig6} emphasizes that such ratios are most diagnostic at earlier epochs.

\section{Conclusion}
\label{conclusion}

We have calculated the line fluxes from $^{56}$Ni and $^{56}$Co decay and escape
for the dominant lines of this beta decay chain, the concomitant hard X-ray Compton and bremsstrahlung continua,
and the iron fluorescence line fluxes for fifteen representative Type Ia explosion models.  
The inclusion of bremsstrahlung (for the first time for most of the fifteen models)
is particularly important for the proper interpretation of the anticipated NuSTAR data.
We have done this without preconceived notions of which explosion models or modalities 
might be preferred from previous considerations.  From the line light curves and ratios, 
continuum band fluxes and their temporal evolution, and rise times and 
peak magnitudes, one may hope to constrain the various physical parameters of the explosion,
and, perhaps, eliminate classes of explosion models while determining explosion parameters.
In principle, the $^{56}$Ni yields and distributions, kinetic energies, and total ejecta masses 
(among other things) can be measured.  Our model results are available in electronic form 
at \url{http://www.astro.princeton.edu/\~{}burrows/sn2014j/} to aid in the interpretation of the
hard photon data anticipated in the ongoing SN 2014J, as well as future, observing campaigns.

\acknowledgments
A.B. acknowledges association with the Joint Institute for Nuclear Astrophysics (JINA, NSF PHY-0822648),
and funding through the NSF PetaApps program, under award OCI-0905046 via a subaward 
no. 44592 from Louisiana State University to Princeton University. We thank P. Hoflich, S. Kumagai, 
K. Nomoto, P. Pinto, and S. Woosley for access to their supernova models. L.-S.T. thanks P.A. Milne for
assistance in compiling the supernova models.

\newpage

\begin{figure}
\begin{center}
\includegraphics[height=.55\textheight,angle=90]{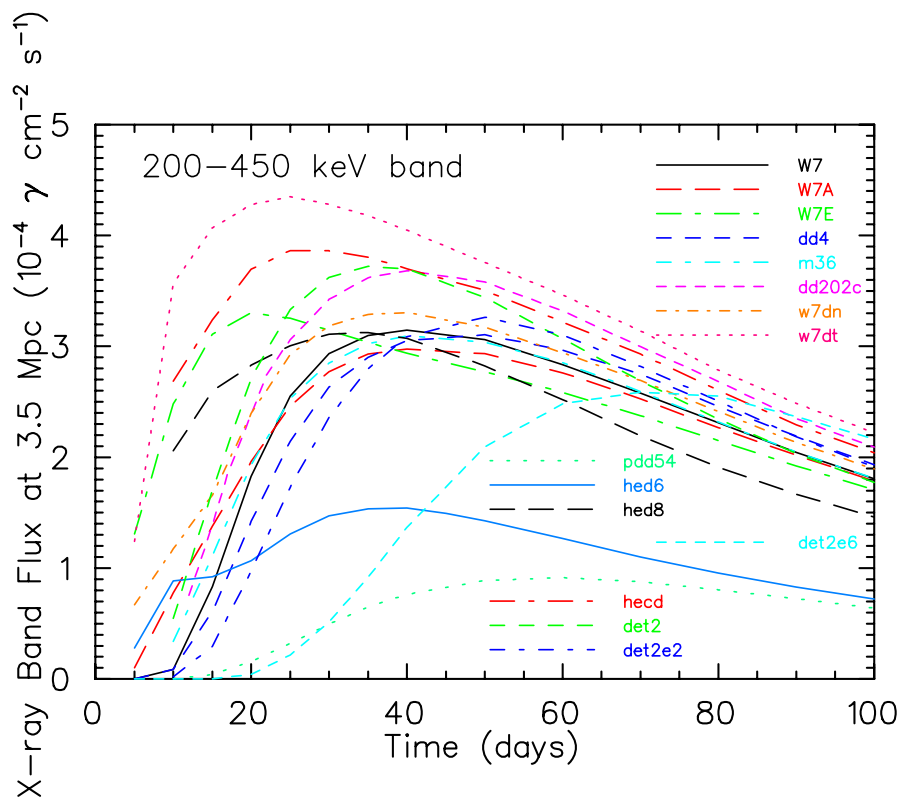}
\end{center}
\caption{The light curves in the 200$-$450 keV band for all fifteen models studied 
in this paper.  Note the diagnostic differences in the evolution of this band flux.
It is expected that the SGD detector on ASTRO-H will be sensitive to many Type Ia
models in this band out to distances slightly greater than $\sim$10 Mpc.
}
\label{fig12b}
\end{figure}

\begin{figure}
\begin{center}
\includegraphics[height=.55\textheight,angle=90]{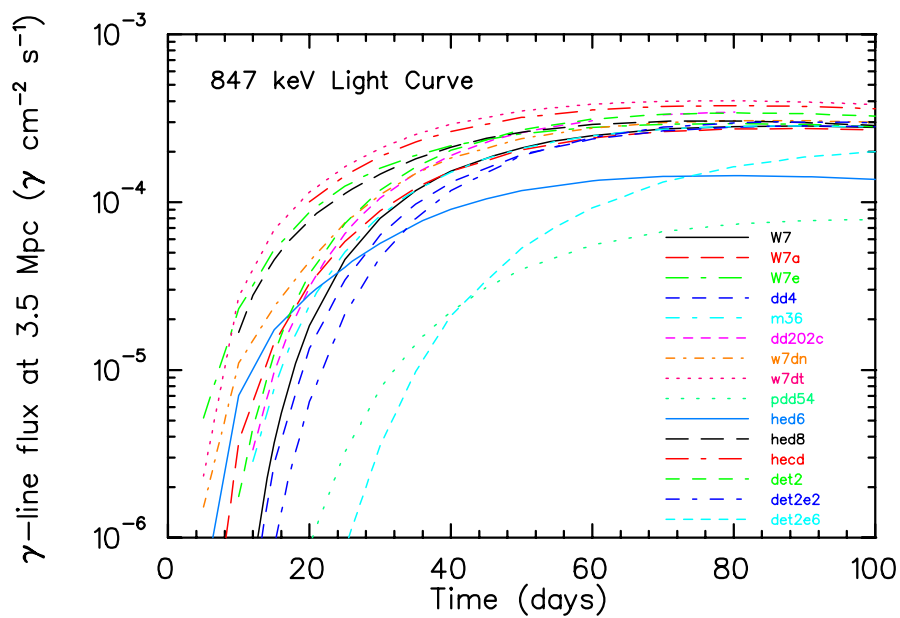}
\end{center}
\caption{The light curves in photons cm$^{-2}$ s$^{-1}$ versus time (in days after explosion)
of the 847 keV line of $^{56}$Co decay at a distance of 3.5 Mpc 
for fifteen Type Ia explosion models in the literature.
The peak emission occurs between day 60 and day 100, but the time at which this light curve 
first exceeds a given threshold is a sharp function of model.  The fluxes at day 40, for instance, vary 
by almost an order of magnitude, and the fluxes at day 20 can vary by a factor of thirty.  
} 
\label{fig3}
\end{figure}

\begin{figure}
\begin{center}
\includegraphics[height=.55\textheight,angle=90]{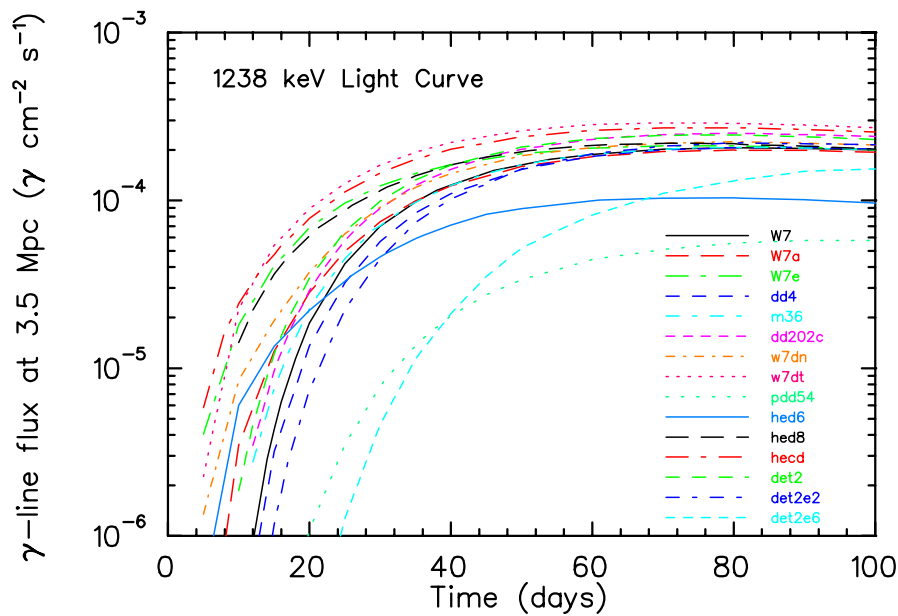}
\end{center}
\caption{The same as Figure \ref{fig3}, but for the 1238 keV line of $^{56}$Co.
} 
\label{fig4}
\end{figure}

\begin{figure}
\begin{center}
\includegraphics[height=.55\textheight,angle=90]{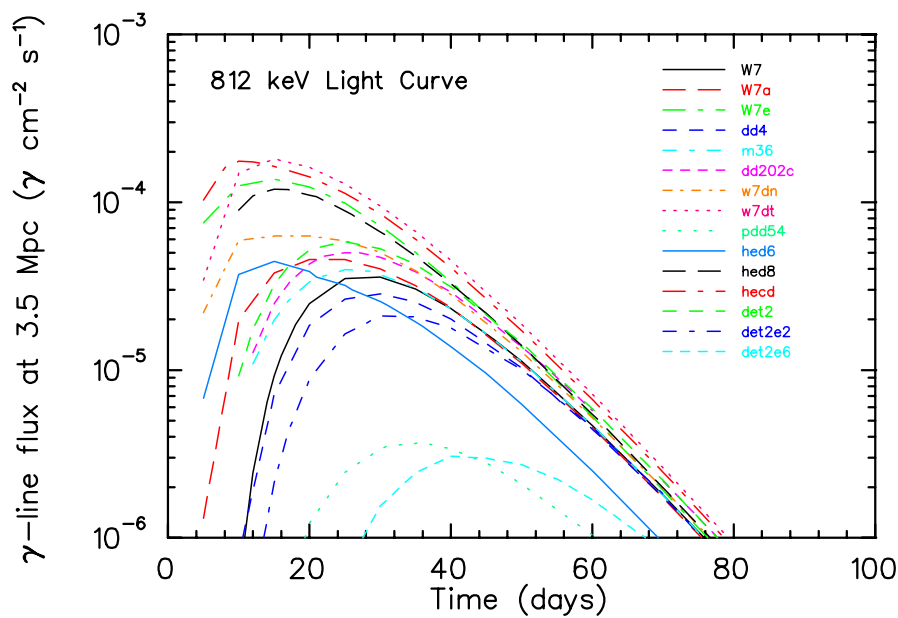}
\end{center}
\caption{The same as Figure \ref{fig3}, but for the 812 keV line of $^{56}$Ni.
Note that the peak values can vary by approximately two orders of magnitude, that the peaks occur earlier
than those for the 847 and 1238 keV lines, and that they decay much faster.  The 812 keV line is one of the most 
diagnostic of model, but is more difficult to see.
} 
\label{fig5}
\end{figure}

\begin{figure}
\begin{center}
\includegraphics[height=.55\textheight,angle=90]{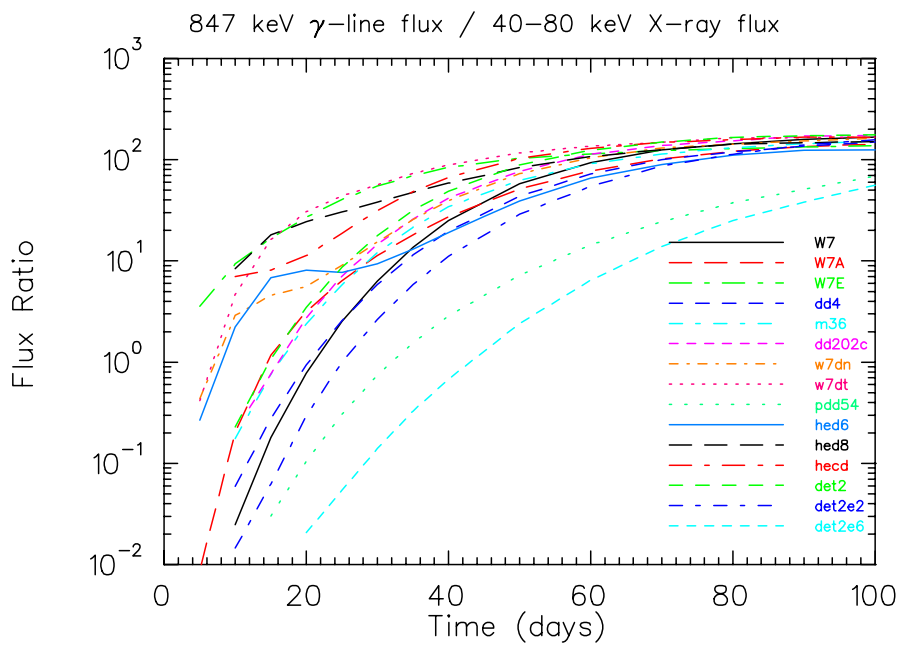}
\end{center}
\caption{The ratio of the 847 keV line flux to the total flux in a 40 to 80 keV bin versus time (in days since explosion), for 
all fifteen models investigated in this study.  Such flux ratios are very diagnostic of model, varying as they do 
at a given epoch after day 20 and from model to model by factors of more than one thousand.  
} 
\label{fig6}
\end{figure}

\end{document}